# Problem of Multiple Diacritics Design for Arabic Script

## Mohamed Hssini[1], Azzeddine Lazrek[2]

[1,2]Department of Computer Science, Caddi Ayad University, Marrakech, Morocco

***Abstract:-*** **This study focuses on the design of multiple Arabic diacritical marks and to developing a model that generates the stacking of multiples Arabic diacritics in order to integrate it into a system of Arabic composition. The problem concerns the presence of multiple diacritics on a single basic letter. This model is based on the layering composition. The combination of diacritics with letters requires a basic layering to combine any diacritics in the word with their base letter, without having to deal individually and separately each pair of base letter and diacritics.**

## I. INTRODUCTION

Multiple Arabic diacritical marks pile up on the basis letter in a manner different from to stacking multiple diacritics in Latin script. They are composed in horizontal layers in order to proper distributing the spaces in the word. Their presence corresponds to a high degree of sophistication and requires complex mechanisms to manage their positions. So, their design is an unmoving challenge for typesetting Arabic. This study focuses on the design of multiple Arabic diacritical marks. Following our studies on the design issues of simple Arabic diacritics, we present here a model that generates the positioning and sizing multiples Arabic diacritics in order to integrate it into a system of text composition.

To address it in this study, we will discuss the following topics: after this introduction, we compare in section 2 the design of multiple Arabic diacritic and there design in print. In section 3, we present the Unicode approach that governs the computing treatment of diacritics, and the rendering of these marks by digital fonts. Section 4 presents our proposed algorithm for rendering multiple Arabic diacritics. The connected problems to an implementation and integration in digital typography and the results make object of section 5. Eventually, we end with conclusions and perspectives.

## II. GENERALITIES

The history of Arabic diacritics offers the potential to frame the historical context in which have appeared strategies for their digital rendering. The graphical filiation from calligraphy to print deserves mention for several reasons. Firstly and before studying numerical methods management Arabic diacritics, we should explore the development of calligraphic design Arabic diacritics and semantic graph, and then examine the historical context that preceded the arrival of typography digital.

The investigation of the transition from calligraphy to typography allows highlight the way in which the practice was viewed the material composition of writing multiple Arabic diacritics. This section focuses on the goal and tries, based on chronological filiations, to clarify the influence of printing by movable type on composition of these brands. Taking this historical approach, it studied changes that the Arabic script has undergone over time, starting by calligraphy as part of the history of Arabic typography.

### 2.1 Arabic diacritical marks
### 21.1.1 Definition

Diacritics are marks that accompany graphically basic letter to transcribe various functions in writing and phonology. They decorate the base letters, a way to change their meaning, indicate their pronunciation, determine their phonetic value and improve the aesthetics of the graphic or word that contains. The word "diacritic" comes from the Greek word "Diacriticos" which means "which distinguishes" [1].

### 21.1.2 Classification

In Arabic script, there are several types of diacritics: the diacritical points, short vowels, diacritics verbalization and various decorative signs.

### 2.1.2.1 Diacritical points

Depending on number of ambiguity, we added some diacritics to invent new letters from simple letters, and to distinguishes graphemes of homographs [1].

**2.1.2.2 Language's diacritics**: differentiate the letter's consonants, are very important for semantic. They appear as:

- **Diacritics above**: placed above a letter, as Fatha, Damma, Sukun.

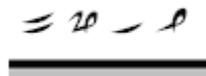

Fig. 1.    Arabic diacritics above

- **Diacritics below**: placed below a letter, as Kasra or Kasratan.

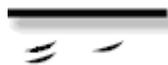

Fig. 2.    Arabic diacritics below

- **Diacritics through**: placed through a letter, as Jarrat Wasl.

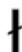

Fig. 3.    Jarrat Wasl through Alef

**2.1.2.3 Diacritic Shadda:** this sign means doubling the base consonant.

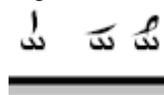

Fig. 4.    Diacritic Shadda

**2.1.2.4 Aesthetic diacritics:** often filled space created when extending some letters, to improve the aesthetic.

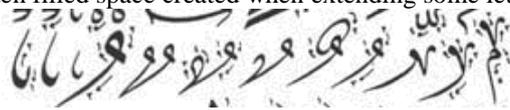

Fig. 5.    Arabic aesthetic diacritics

**2.1.2.5 Explanatory diacritics**: positioned to distinguish letters with dots and letters without dots.

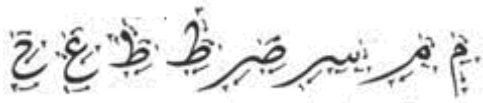

Fig. 6.    Arabic explanatory diacritics

**2.1.3 Multiple Arabic diacritics**
**2.1.4.1 Presence**
The presence of multiple diacritics on a basic letter occurs when some signs with the same or different types are combined one the basic letter.
**2.1.4.2 Stacking diacritics**
The Arabic writing with diacritics is a series of overlapping layers horizontally. The allographs, points, vowels, signs of verbalization and decorative signs are added in successive layers [2].

Visual conventions of other brands remain silent. They can balance the image aesthetically and distributing spaces. These phases are as follows:

- Choising allographs without dots or diacritics, which are a series of words nested;
- Positioning points;
- Adding Shadda sign;
- Adding marks of vocalization;
- Adding marks of verbalization;
- Inserting explanatory diacritics;
- Adding aesthetic signs.

The size of some Arabic diacritics varies in order to fill the void, for a harmonious enjoyment and composition of Arabic script. The use of the relative size of the elements against each others can focus attention to a focal point [1].

The variations in shape, size and position can all play important grammatological roles. The application of these features is subsequently explored in relation to the practice of the Arabic script. Arabic tradition uses the differences in size and position to define the information contained in the graph marks. These practices built in writing that the design of the diacritic is multidimensional space.

The position of a diacritic is a function of its size and dimensions of the base letter and letters and their neighboring signs.

## 2.2 Arabic diacritical marks and print

The typographic approach, which results from the Gutenberg model, is to represent writing by segmenting it in individual typographic signs. The constituent unit was the metal block [3].

Generally, the technique of printing was not able to offer the possibility to control the graphic aspects of the Arabic text with extra level of authenticity: adapting grapheme of each Arabic letter to constraints imposed by the support, controlling the body and the position of allograph into the cursive attachment and justification, managing the contexts for freely associate letters and diacritics, and so on.

Above and below the basic line of writing, diacritics and decorative signs modify and extend the interline spacing. The introduction of decorative signs could break the line even further, jump through the levels of writing or radically changing direction to preserve the aesthetic balance. The design of diacritical text comes into spatial relations and position with the surrounding texts and space [4]. It is constructed according to the available space and as a function of the amount of content. On solution is to redesign all diacritics in one mark.

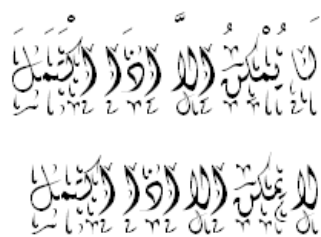

Fig. 7.      Redesigning multiple diacritics in one mark

## III.  RENDERING OF MULTIPLE ARABIC DIACRITICS

Support of these diacritics requires a number of features. Several levels of support are possible. The first challenge was coding, especially in the early days of the computer with its technical limitations.

### 3.1 CODING

In Unicode, the number of diacritical marks that may accompany a base character is not limited. If combining characters can affect the same typographical area, order of coded characters determines the order of display [6]. Diacritics are willing to turn from the glyph corresponding to the base character and then move away. Combining characters placed above a base character stack vertically, starting with the first in the order of logical storage [7] [8][9]. We continue to the top stacking all the signs of the need for the number of character codes which follow the base character. The situation is reversed for combining characters under a base character: characters combinatorial stack down. We call this order centrifugal order.

### 3.2 SMART FONT

Many writing systems around the world have complex rules governing the way the elements of the script are written. These kinds of writing systems require smart font technology to be rendered properly on the computer. Among these challenges, there is stacking diacritics. Digital fonts treat this problem following many methods  [11]:

•   Make one glyph containing all glyphs diacritics;
•   Leave a diacritic free and represent others with the base glyph in a single glyph;
•   Represent only one among other diacritical sign is Shadda;

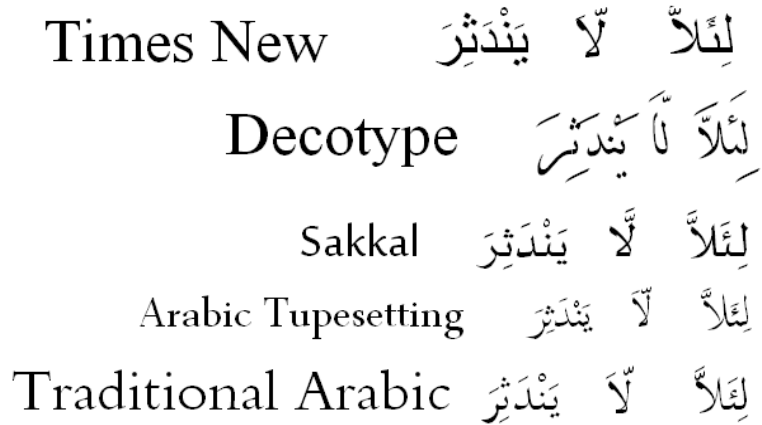

Fig. 8.     Rendering multiple Arabic diacritics in some digital fonts

It is therefore evident that the adjustment diacritics by pairs glyphs has its limits facing a complete and authentic diacritization of Arabic text. In Arabic script, kerning and adjustment by pairs of glyphs are not sufficient. Often more diacritics are combined with the basic consonant and with various positions and various sizes. An interesting detail is the inability of the existing Arabic typography to manage diacritization with all its complexities with the choice of allographs and ligatures composition etc. at the same time: one cancels the other one. However, the sophisticated script with ligatures is usually associated to complete diacritization. To cope with this problem, it is carried to review the level of atomicity treatment of Arabic diacritics and try to approach a process model of diacritization in Arabic calligraphy. We have seen that the methods of combining diacritical marks can be summarized in four types: ligatures, kerning, Anchor points and substitution. In 1) we have presented our attempt to improve the rendering of Arabic diacritics at the cast and show the limitations of these methods when managing multiple Arabic diacritics.

## IV.   PROPOSED ALGORITHM

### 4.1 MODELING STEPS
The construction of a computer model that generates this process diacritization Arabic calligraphy led to the development of a model that generates the production of Arabic script: it brings us then to understand and formalize the new phenomena such as contextuality or dependence between ingredients textual phenomena are also present, although sometimes in a disguised manner in manuscripts historical and even current models. The calligraphic diacritization is a process that consists of a set of related activities and is divided into stages. These activities transform inputs into outputs. Each step consists of tasks and generates its own data. Positioning and sizing of the diacritics are the results of a series of situations of managing contexts.

#### 4.1.1 UNIT OF PROCESSING
In Arabic calligraphy, the word is treated as a coalesced mass in the management of diacritics. The calligraphic word with diacritics was born from a combination of elements all arranged to form a coherent and unified set.

#### 4.1.2 MODELING CONTEXTS
To extract rules management Arabic diacritics, it is sufficient to analyze contexts which may appear in a position or size a diacritical mark. We face two problems related to the identification of contexts and the volume of data available for analysis.

### 4.2 ALGORITHM
To adapt the process of diacritization in Arabic calligraphy to typographic implementation, we modify the proposed algorithm in [2] as follows:
- At the input, take a diacritical word as a chain of $l_1 d_{1,1} \dots d_{1,p1}$ $l_2 d_{2,1} \dots d_{2,p2} \dots l_N d_{N,1} \dots d_{N,pN}$ where: $l_i$ mean the basic letters and $d_i$ design diacritics;
- Subdivide $l_1 d_{1,1} \dots d_{1,p1}$ $l_2 d_{2,1} \dots d_{2,p2} \dots l_N d_{N,1} \dots d_{N,pN}$ chain into two chains: $l_1 l_2 \dots l_N$ and $d_{1,1} \dots d_{1,p1}$ $d_{2,1} \dots d_{2,p2} \dots d_{N,1} \dots d_{N,pN}$
- Enter the word without diacritics $l_1 l_2 \dots l_N$;
- Sort $d_{1,1} \dots d_{1,pN}$ $d_{2,1} \dots d_{2,p2} \dots d_{N,1} \dots d_{N,pN}$ diacritics according to their natures;
- Compos layers representing classes;

- Choose the size of each diacritic adjust sizes and positions in $d_{1,1}…d_{1,p1}$ $d_{2,1}…d_{2,p2}$ … $d_{N,1}$ … $d_{N,pN}$ chain of diacritics according to the calligraphic rules;
- At the output, display the word with diacritics following calligraphic rules.

## V.  PRACTICE RESULTS

### 5.1 IMPLEMENTATION ENVIRONMENT

The implementation of the model of layer composition confronts the various technological constraints. It will discuss the infrastructure to prepare to examine the applicability of the model to the composition systems and its compatibility with the existing text models. The integration of our model of layered compositions requires the establishment of a solid infrastructure. Decreases experienced ascendants and descendants of the Arabic alphabet, due to problems of rationality influence on dynamited sizes and positions of Arabic diacritics decrease because of surface empty complete.

Arabi system [16] is a set of packages for typesetting Arabic text in $L^A T_E X$,. It allows the possibility of using Arabic calligraphy along the line in Latin environment. Arabi use digital Arabic fonts in Metafont, TrueType, and PostScript. Arabi system is characterized by:

- Uses all $T_E X$ engine supporting right to left direction;
- Analysis contexts in the digital line and not in the system;
- Its flexibility and coexistence with other packages.

Must be the provision of infrastructure in order to integrate the results obtained in the Arabi system, and are as follows:

- Develop package includes heuristics engine and you masterminding regulations strings.
- Use package ifthen in order to provide Macros the assessment and comparison.
- This package is a set of software macros enable management regulations and at the same time use in programming.
- classification of glyphs in a font by length, width and depth.

### 5.2 RESULTS

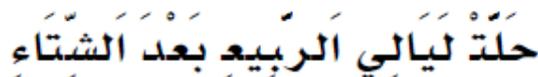

Fig. 9.    Initial rendering of Arabi system

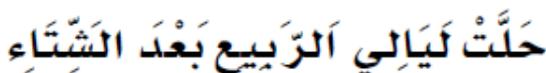

Fig. 10.    Modified rendering of Arabi system

The model was corrected rendering multiple diacritics in the Arab system composition Arabi.

## VI.  CONCLUSION

This work focuses on the development of a model that generates the stacking multiples Arabic diacritics in order to integrate it into a text composition system. The authenticity of such desired record is needed to bear in mind the rules governing the design of calligraphic marks. The integration of the model with the rules retrieved in a composition system text requires the revision of the infrastructure. The proposed model has helped improve the rendering of these marks in the Arab system composition. However, three main criticisms may be intended for our work: the size of the corpus, the rules and the number of contexts to be managing and the preparation of the typographic infrastructure. This work has numerous perspectives, among which we can mention: textual modeling, managing textual contexts and recognition of the calligraphic style from the morphology of Arabic word.

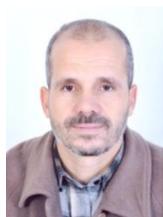

**Mohamed Hssini** is a Ph.D. student in Department of Computer Science at Cadi Ayyad University. He is a member of *multilingual scientific e-document processing* team. His current main research interest is multilingual typography, especially the publishing of Arabic e-documents while observing the strict rules of Arabic calligraphy.

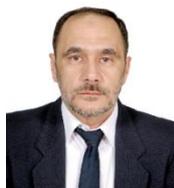

**Azzeddine Lazrek** is full Professor in *Computer Science* at Cadi Ayyad University in Marrakesh. He holds a *PhD* degree in *Computer Science* from Lorraine Polytechnic National Institute in France at 1988, in addition to a *State Doctorate* Morocco at 2002. Prof. Lazrek works on *Communication multilingual multimedia e-documents*. His areas of interest include: multimedia information processing and its applications, particularly, electronic publishing, digital typography, Arabic processing and the history of science. He is in charge of the research team *Information Systems and Communication Networks* and the research group *Multilingual scientific e-document processing*. He is an *Invited Expert* at W3C. He leads a *multilingual e-document composition* project with other international organizations.